\documentstyle[twoside,fleqn,espcrc2,epsf]{article}

\newcommand{\AmS}{{\protect\the\textfont2
  A\kern-.1667em\lower.5ex\hbox{M}\kern-.125emS}}

\title{
\vspace*{-35pt}
{\normalsize \hfill {\sf UTCCP-P-73}} \\
\vspace*{-6pt}
{\normalsize \hfill {\sf Sept.\ 1999}} \\
Eta meson mass and topology in QCD with two light flavors\thanks{talk
presented by R.~Burkhalter at Lattice '99}}
 
\author{
{CP-PACS Collaboration}:
A.~Ali~Khan\address{Center for Computational Physics,
University of Tsukuba, Tsukuba, Ibaraki 305-8577, Japan}, 
S.~Aoki\address{Institute of Physics,
University of Tsukuba, Tsukuba, Ibaraki 305-8571, Japan},
R.~Burkhalter$^{\rm a,b}$, 
S.~Ejiri$^{\rm a}$, 
M.~Fukugita\address{Institute for Cosmic Ray Research, 
University of Tokyo, Tanashi, Tokyo 188-8502, Japan}, 
S.~Hashimoto\address{High Energy Accelerator Research Organization (KEK), 
Tsukuba, Ibaraki 305-0801, Japan}, 
N.~Ishizuka$^{\rm a,b}$,
Y.~Iwasaki$^{\rm a,b}$, 
K.~Kanaya$^{\rm a,b}$, 
T.~Kaneko$^{\rm a}$, 
Y.~Kuramashi$^{\rm d}$,
T.~Manke$^{\rm a}$,
K.~Nagai$^{\rm a}$,
M.~Okawa$^{\rm d}$, 
H.P.~Shanahan\address{DAMTP, 21 Silver St., University of Cambridge,
Cambridge, CB3 9EW, England, U.K.},
A.~Ukawa$^{\rm a,b}$ and T.~Yoshi\'e$^{\rm a,b}$
}

\begin{document}

\begin{abstract}

We present results for the mass of the flavor singlet meson calculated 
on two-flavor full QCD configurations generated by 
the CP-PACS full QCD project.
We also investigate topological charge fluctuations 
and their dependence on the sea 
quark mass. 

\end{abstract}

\maketitle

\section{INTRODUCTION}

Recently considerable progress has been made in the simulation of full
QCD~\cite{ref:mawhinney}. In particular, sea quark effects 
have been found to lead to a closer agreement of the light
meson spectrum with experiment~\cite{ref:review98,ref:kaneko}. 

Missing from the calculated spectrum, however, has been 
the flavor singlet meson $\eta'$.  
Due to the difficulty of the determination of the
disconnected contribution,  
only preliminary lattice results have been 
available~\cite{ref:Itoh,ref:Kuramashi,ref:Venkataraman}.

In the first half of this article, 
we present new results on this problem. 
Since these are obtained with two flavors of
dynamical quark, we call the flavor singlet meson as $\eta$ and
reserve the name $\eta'$ for the case of $N_f\!=\!3$.

The $\eta'$ meson is expected to obtain
a large mass through the connection to instantons. This leads us to an
investigation of topology in full QCD, presented in the latter half of this
article.

\begin{table}[ht]
\setlength{\tabcolsep}{.2pc}
\newlength{\digitwidth} \settowidth{\digitwidth}{\rm 0}
\catcode`?=\active \def?{\kern\digitwidth}
\caption{Parameters of lattices used in this calculation.}
\label{tab:runs} 
\begin{center}
\begin{tabular}{ccccc}\hline
$N_s^3\!\times\!N_t$ & $\beta$ & $N_s a$ [\mbox{fm}] & $a^{-1}$ [\mbox{GeV}]& 
$m_{\rm PS}/m_{\rm V}$ \\ 
\hline
$12^3\!\times\!24$ & 1.8  & 2.58(3) & 0.92(1) & 0.81--0.55 \\ 
$16^3\!\times\!32$ & 1.95 & 2.45(3) & 1.29(2) & 0.80--0.59 \\ 
$24^3\!\times\!48$ & 2.1  & 2.60(5) & 1.82(3) & 0.80--0.58 \\ 
$24^3\!\times\!48$ & 2.2  & 2.06(6) & 2.30(7) & 0.80--0.63 \\ 
\hline 
\end{tabular}

\end{center}
\vspace{-11mm}
\end{table}

Calculations have been performed on configurations of the CP-PACS full QCD
project~\cite{ref:review98,ref:kaneko}. These have been generated using
an RG-improved gauge action and a tadpole-improved SW clover quark action
at four different lattice spacings and four values of the sea quark mass
corresponding to $m_{\rm PS}/m_{\rm V}\!\sim\!0.8$--0.6. An overview of the
simulation parameters is given in Table~\ref{tab:runs}. More details can be
found in~\cite{ref:review98,ref:kaneko}.

\section{FLAVOR SINGLET MESON}

The mass difference $\Delta m$ between the flavor singlet ($\eta$) and
non-singlet ($\pi$) meson can be extracted from the ratio
\begin{equation}
R(t) = \frac{\langle\eta(t)\eta(0)\rangle_{\rm disc}}
            {\langle\eta(t)\eta(0)\rangle_{\rm conn}}
       \rightarrow 1 - B\exp(-\Delta mt),
\label{eq:ratio}
\end{equation}
where the right hand side indicates the expected behavior at large time
separation $t$.   

In this work, the connected propagator was calculated with the standard
method. For the disconnected propagator we used two methods. 
In the first instance, it was calculated using a volume source
without gauge fixing (the Kuramashi method)\cite{ref:Kuramashi}.  
This measurement was made after every trajectory in the
course of configuration generation for all runs listed in
Table~\ref{tab:runs}.
As for the second method, we employed a U(1) volume noise source with 10 random
noise ensembles for each color and spin combination. This 
was performed only at $\beta=1.95$ on stored configurations separated 
by 10 HMC trajectories.

\begin{figure}[tb]
\vspace{-25pt}
\centerline{\epsfysize=6cm \epsfbox{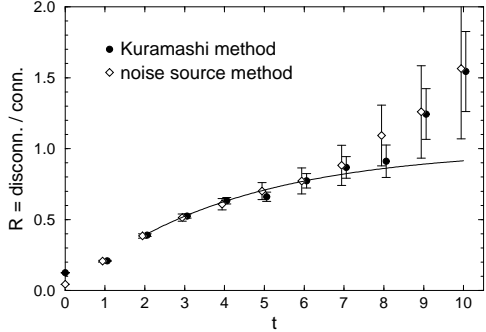}}
\vspace{-40pt}
\caption{Ratio for $\beta=1.95$ and $m_{\pi}/m_{\rho}=0.59$. The solid line 
is from a fit with Eq.~\ref{eq:ratio}.} 
\label{fig:ratio}
\vspace{-20pt}
\end{figure}

Figure~\ref{fig:ratio} compares the ratio $R(t)$ for the two methods. We
observe that they are consistent with each other but that the error is
smaller for the first method. This might be due to the fact that there are
10 times more measurements with it, although binning is made over 50 HMC
trajectories for both cases to take into account auto-correlations. In the
following we only use data obtained with the first method.

\begin{figure}[tb]
\vspace{-25pt}
\centerline{\epsfysize=6cm \epsfbox{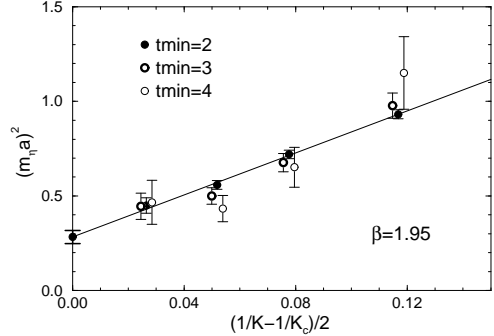}}
\vspace{-40pt}
\caption{Chiral extrapolation of $\eta$ meson mass.}
\label{fig:chireta}
\vspace{-20pt}
\end{figure}

In Fig.~\ref{fig:ratio} we also see that the error of $R(t)$ increases
exponentially, which make the determination of $\Delta m$ via
Eq.~\ref{eq:ratio} impossible at large time separations. The data, however,
shows the expected behavior already beginning from small $t$ and a fit with
Eq.~\ref{eq:ratio} is possible from $t_{min}\!=\!2$.
Increasing $t_{min}$ leads to stable results, as can be seen in
Fig.~\ref{fig:chireta}.

The chiral extrapolation of $m_{\eta}^2$ linear in the quark mass is shown
in Fig.~\ref{fig:chireta}. Contrary to the pion mass, the flavor singlet
meson remains massive in the chiral limit.

\begin{figure}[tb]
\vspace{-25pt}
\centerline{\epsfysize=6cm \epsfbox{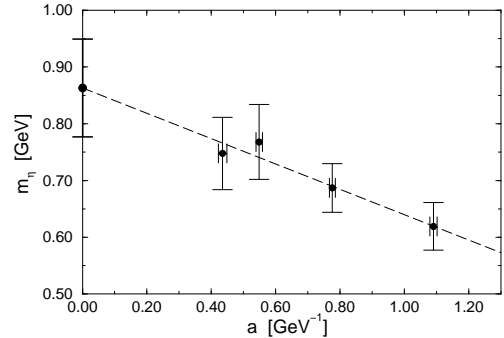}}
\vspace{-40pt}
\caption{Continuum extrapolation of $\eta$ mass.}
\label{fig:eta}
\vspace{-20pt}
\end{figure}

Figure~\ref{fig:eta} shows the $\eta$ meson mass at all measured lattice
spacings after the chiral extrapolation. The scale is set using the $\rho$
meson mass. A linear extrapolation to the continuum limit gives
$m_\eta=863(86)$~MeV. This value lies between the experimental $\eta$(547) 
and $\eta'$(958) masses. We emphasize that a proper comparison with
experiment requires the introduction of a third (strange) quark and a
mixing analysis.

\section{TOPOLOGY}

Studies of topology on the lattice have encountered several difficulties. 
In addition
to the ambiguity of defining a lattice topological charge, 
it was found that topological modes have a very long
auto-correlation time in the case of full QCD with the Kogut-Sussind 
quark action \cite{ref:topo-full}.

We employ the field theoretic definition of the topological charge together
with cooling. For the charge we use a tree-level improved definition which
includes a $1\!\times\!2$-plaquette, hence the $O(a^2)$ terms are removed
for instanton configurations.  For the cooling we compare two choices of
improved actions, both including a $1\!\times\!2$-plaquette term: 1)~a
tree-level Symanzik improved (LW) action and 2)~the RG improved Iwasaki
action.

\begin{figure}[tb]
\vspace{-25pt}
\centerline{\epsfysize=6cm \epsfbox{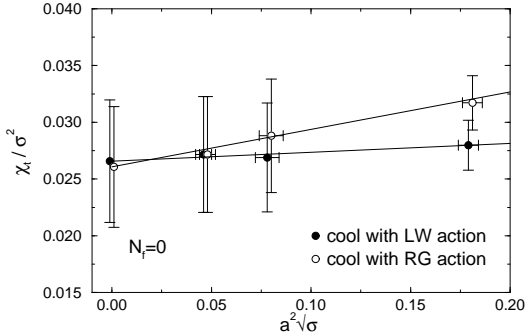}}
\vspace{-40pt}
\caption{Topological susceptibility for $N_f=0$.}
\label{fig:suscCont}
\vspace{-22pt}
\end{figure}

Using different actions for cooling can lead to different values of the
topological charge. This ambiguity is only expected to vanish when the
lattice is fine enough. We have tested this explicitly by simulating the pure
SU(3) gauge theory at three lattice spacings in the range $a \sim
0.2$--0.1~fm and at a constant size of 1.5~fm. As Fig.~\ref{fig:suscCont}
shows, the topological susceptibilities $\chi_t\!=\!\langle Q^2\rangle/V$
for the two cooling actions converge to a common value towards the
continuum limit. Using $\sqrt{\sigma}\!=\!440$~MeV we obtain
$\chi_t\!=\!(178(9)\,{\rm MeV})^4$, in agreement with previous
studies~\cite{ref:topo-pure}.

\begin{figure}[b]
\vspace{-45pt}
\centerline{\epsfysize=5.55cm \epsfbox{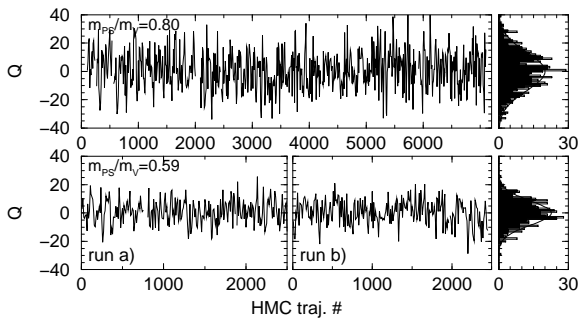}}
\vspace{-45pt}
\caption{Charge history at $\beta=1.95$.}
\label{fig:history}
\vspace{-2pt}
\end{figure}

In full QCD we have so far measured the topological charge at
$\beta\!=\!1.95$. Figure~\ref{fig:history} shows the time history for two
quark masses. Auto-correlation times are visibly small even for the
smallest quark mass. For the Wilson quark action rather short
auto-correlation times have been reported in Ref.~\cite{ref:SESAM}. The
fact that we find even shorter auto-correlations might be explained by the
coarseness of our lattice.

\begin{figure}[tb]
\vspace{-25pt}
\centerline{\epsfysize=6cm \epsfbox{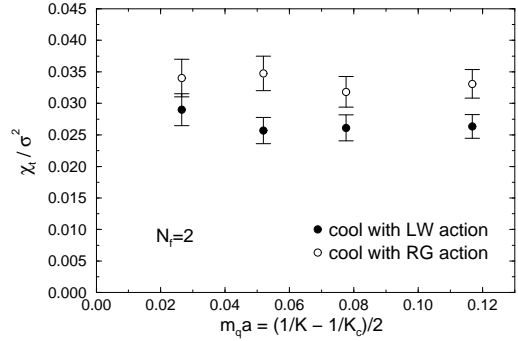}}
\vspace{-40pt}
\caption{Topological susceptibility at $\beta=1.95$.}
\label{fig:suscept}
\vspace{-22pt}
\end{figure}

Based on the anomalous flavor-singlet axial vector current Ward identity,
one expects the topological susceptibility to vanish in the chiral
limit. Indeed, Fig.~\ref{fig:history} shows the width to be shrinking with
the quark mass. The decrease, however, is not sufficient; 
as we find in Fig.~\ref{fig:suscept}, the 
dimensionless ratio $\chi_t/\sigma^2$, with $\sigma$ calculated for each 
sea quark mass, does not vary much with the quark mass, 
and takes a value similar to that for pure gauge theory.  
To understand the origin of this
behavior, more investigations at different lattice spacings will be needed.

\vspace{1.3mm}
This work is supported in part by Grants-in-Aid 
of~the~Ministry~of~Education~(Nos.~09304029,
10640246,~10640248,~10740107,~11640250, 11640294, 11740162). 
SE and KN are JSPS Research Fellows.  AAK, TM and HPS are
supported by JSPS Research for the Future Program. HPS is supported by
the Leverhulme foundation.


\vspace{-6pt}
 
\begin{thebibliography}{9}

\vspace{-3pt}

\bibitem{ref:mawhinney}
R.~Mawhinney, review in these proceedings.

\bibitem{ref:review98}
R.~Burkhalter,
Nucl.\ Phys.\  {\bf B} (Proc.Suppl) {\bf 73} (1999) 3.

\bibitem{ref:kaneko}
T.~Kaneko {\it et al.}, 
CP-PACS collaboration,
these proceedings.

\bibitem{ref:Itoh}
S.~Itoh {\it et al.}, 
Phys.\ Rev.\ {\bf D36} (1987) 527.

\bibitem{ref:Kuramashi}
Y.~Kuramashi {\it et al.}, 
Phys.\ Rev.\ Lett.\ {\bf 72} (1994) 3448.

\bibitem{ref:Venkataraman}
L.~Venkataraman and G.~Kilcup, 
hep-lat/9711006.

\bibitem{ref:topo-full}
Y.~Kuramashi {\it et al.}, 
Phys.\ Lett.\ {\bf B313} (1993) 425;
B.~All\'{e}s {\it et al.}, 
Phys.\ Lett.\ {\bf B389} (1996) 107.

\bibitem{ref:topo-pure}
For a recent review see 
J.~Negele, Nucl.\ Phys.\  {\bf B} (Proc.Suppl) {\bf 73} (1999) 92.

\bibitem{ref:SESAM}
B.~All\'{e}s {\it et al.}, 
Phys.\ Rev.\ {\bf D58} (1998) 071503.


\end{thebibliography}
\end{document}